\documentclass[12pt]{article}
\usepackage{graphicx,url,apalike}

\newcommand{\ra}{\rightarrow}
\newcommand{\kiw}{\ensuremath{\delta_{i}(w)}}
\newcommand{\kiwo}{\ensuremath{\delta_{i}(w,\vec{a}_{-i})}}
\newcommand{\kijw}{\ensuremath{\delta_{ij}(w)}}
\newcommand{\kijwo}{\ensuremath{\delta_{ij}(w,\vec{a}_{-j})}}
\newcommand{\kijkw}{\ensuremath{\delta_{ijk}(w)}}

\newcommand{\kiwa}{\ensuremath{\delta_{i}:w \ra a_{i}}}
\newcommand{\kiwoa}{\ensuremath{\delta_{i}: (w,\vec{a}_{-i}) \ra a_{i}}}
\newcommand{\kijwa}{\ensuremath{\delta_{ij}: w \ra a_{j}}}
\newcommand{\kijwoa}{\ensuremath{\delta_{ij} : (w,\vec{a}_{-j}) \ra a_{j}}}
\newcommand{\kijkwa}{\ensuremath{\delta_{ijk}: w \ra a_{k}}}

\newcommand{\chart}[1]{\centerline{\includegraphics*[width=3in,height=2in]{#1}}}
\newcommand{\chartl}[1]{\centerline{\includegraphics*[width=3.25in,height=2in]{#1}}}
\newcommand{\chartll}[1]{\centerline{\includegraphics*[width=4.5in,height=2.5in]{#1}}}

\begin{document}
\newtheorem{theorem}{Theorem} 
\newtheorem{corollary}{Corollary}

\title{Learning Nested Agent Models in an Information Economy}
\author{Jos\'{e} M. Vidal and Edmund H. Durfee \\
  Artificial Intelligence Laboratory  \\
  University of Michigan \\
  1101 Beal Avenue \\ Ann Arbor, MI 48109-2110 \\ 
  jmvidal@engr.sc.edu, durfee@umich.edu }
\date{}
\maketitle
\begin{abstract}
  We present our approach to the problem of how an agent, within an
  economic Multi-Agent System, can determine when it should behave
  strategically (i.e. learn and use models of other agents), and when
  it should act as a simple price-taker. We provide a framework for
  the incremental implementation of modeling capabilities in agents,
  and a description of the forms of knowledge required. The agents
  were implemented and different populations simulated in order to
  learn more about their behavior and the merits of using and learning
  agent models. Our results show, among other lessons, how savvy
  buyers can avoid being ``cheated'' by sellers, how price volatility
  can be used to quantitatively predict the benefits of deeper models,
  and how specific types of agent populations influence system
  behavior.
  
\end{abstract}
\section{Introduction}

In open, multi-agent systems, agents can come and go without any
central control or guidance, and thus how and which agents interact
with each other will change dynamically. Agents might try to
manipulate the interactions to their individual benefits, at the cost
of the global efficiency.  To avoid this, the protocols and mechanisms
that the agents engage in might be constructed to make manipulation
irrational \cite{rules:of:encounter}, but unfortunately this strategy
is only applicable in restricted domains.  By situating agents in an
economic society, as we do in the University of Michigan Digital
Library (UMDL), we can make each agent responsible for making its own
decisions about when to buy/sell and who to do business with
\cite{atkins:96}.  A market-based infrastructure, built around
computational auction agents, serves to discourage agents from
engaging in strategic reasoning to manipulate the system by keeping
the competitive pressures high.  However, since many instances can
arise where imperfections in competition could be exploited, agents
might benefit from strategic reasoning, either by manipulating the
system or not allowing others to manipulate them.  But strategic
reasoning requires effort.  An agent in an information economy like
the UMDL must therefore be capable of strategic reasoning and of
determining when it is worthwhile to invest in strategic reasoning
rather than letting its welfare rest in the hands of the market
mechanism.

In this paper, we present our approach to the problem of how an agent,
within an economic MAS, can determine when it should behave
strategically, and when it should act as a simple price-taker. More
specifically, we let the agent's strategy consist of learning nested
models of the other agents, so the decision it must make refers to
which of the models will give it greater gains.  We show how, in some
circumstances, agents benefit by learning and using models of others,
while at other times the extra effort is wasted.  Our results point to
metrics that can be used to make quantitative predictions as to the
benefits obtained from learning and using deeper models.

\subsection{Related work}
\label{sec:related}

Different research communities have run across the problems that arise
from having agents learning in societies of learning agents. The
studies of \cite{shoham:92}, and \cite{glance:93} focus on very simple
but numerous agents and emphasize their emergent behavior.
\cite{hu:96} show that learning agents in an economic domain sometimes
converge to globally sub-optimal equilibria. The work on agent-based
modeling \cite{hubler,axelrod:84,artificial:societies} of complex
systems studies slightly more complicated agents that are meant as
stand-ins for real world agents (e.g.  insects, communities,
corporations, people).

All these researchers used agents whose learning abilities consist of
choosing from among a set of fixed strategies. They do not explicitly
consider the fact that the agents inhabit communities of learning
agents. That is, their agents do not try to model other agents. We
address this issue and try to determine when and which models an agent
should keep.

Within the MAS community, some work \cite{ACL:96} has focused on how
artificial AI-based learning agents would fare in communities of
similar agents. For example, \cite{nadella:97} and \cite{terabe:97}
show how agents can learn the capabilities of others via repeated
interactions, but these agents do not learn to predict what actions
other might take. Most of the work in MAS also fails to recognize the
possible gains from using explicit agent models to predict agent
actions. \cite{tambe:96} is an exception and gives another approach
for using nested agent models. However, they do not go so far as to
try to quantify the advantages of their nested models or show how
these could be learned via observations. We believe that our research
will bring to the foreground some of the common observations seen in
these research areas and help to clarify the implications and utility
of learning and using nested agent models.

\nocite{vidal:96c}

\section{Description of the UMDL}

The UMDL project is a large-scale, multidisciplinary effort to design
and build a flexible, scalable infrastructure for rendering library
services in a digital networked environment. In order to meet these
goals, we chose to implement the library as a collection of
interacting agents, each specialized to perform a particular task.
These agents buy and sell goods/services from each other, within an
artificial economy, in an effort to make a profit. Since the UMDL is
an open system, which will allow third parties to build and integrate
their own agents into the architecture, we treat all agents as purely
selfish.

\subsection{Implications of the information economy.}
Information goods/services, like those provided in the UMDL, are very
hard to compartmentalize into equivalence classes that all agents can
agree on. For example, if a web search engine service is defined as a
good, then all agents providing web search services can be considered
as selling the same good. It is likely, however, that a buyer of this
good might decide that seller $s1$ provides better answers than seller
$s2$. We cannot possibly hope to enumerate the set of reasons an agent
might have for preferring one set of answers (and thus one search
agent) over another, and we should not try to do so. It should be up
to the individual buyers to decide what items belong to the same good
category, each buyer clustering items in possibly different ways.
 
This situation is even more evident when we consider an information
economy rooted in some information delivery infrastructure (e.g. the
Internet). There are two main characteristics that set this economy
apart from a traditional economy.

\begin{itemize}
  
\item There is virtually no cost of reproduction. Once the information
  is created it can be duplicated virtually for free.
  
\item All agents have virtually direct and free access to all other
  agents.

\end{itemize}

If these two characteristics are present in an economy, it is useless
to talk about supply and demand, since supply is practically infinite
for any particular good and available everywhere. The only way agents
can survive in such an economy is by providing value-added services
that are tailored to meet their customers' needs. Each provider will
try to differentiate his goods from everyone else's while each buyer
will try to find those suppliers that best meet her value function. We
propose to build agents that can achieve these goals by learning
models of other agents and making strategic decisions based on these
models. These techniques can also be applied, with variable levels of
efficacy, to traditional economies.

\section{A Simplified Model of the UMDL}

In order to capture the main characteristics of the UMDL, and to
facilitate the development and testing of agents, we have defined an
``abstract'' economic model.  We define an economic society of agents
as one where each agent is either a {\em buyer} or a {\em seller} of
some particular good. The set of buyers is $B$ and the set of sellers
is $S$.  These agents exchange goods by paying some price $p \in P$,
where $P$ is a finite set. The buyers are capable of assessing the
quality of a good received and giving it some value $q \in Q$, where
$Q$ is also a finite set.

\begin{figure}[p]
\begin{center}
\leavevmode 
\chartl{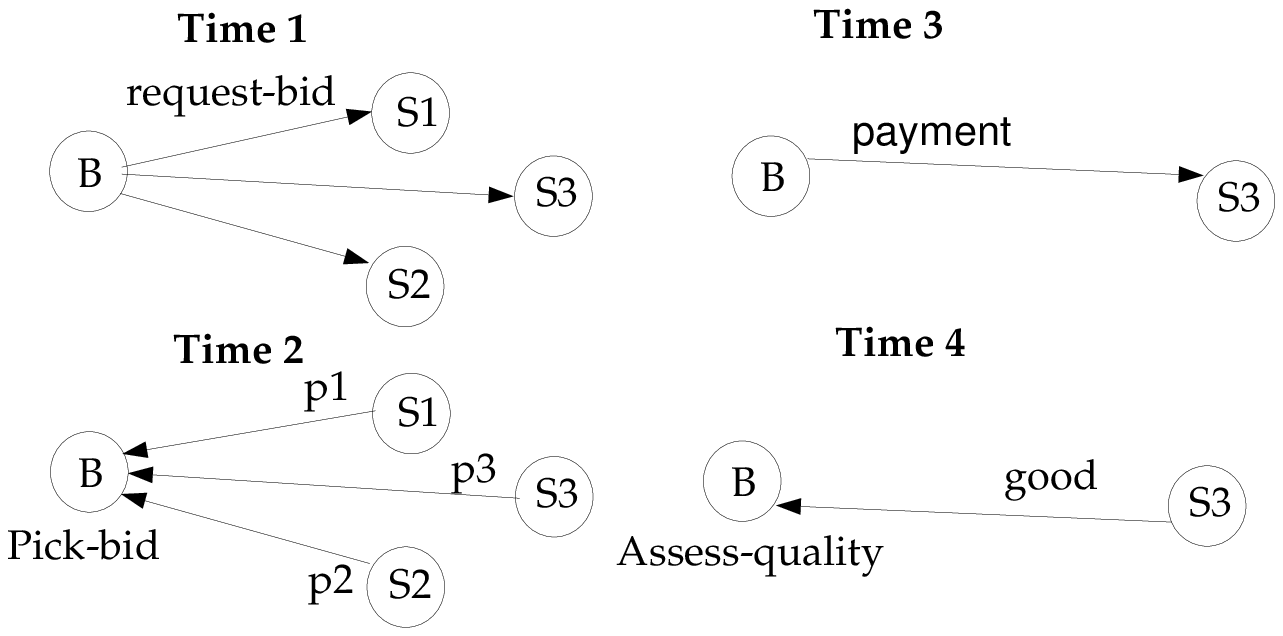}
\caption{View of the protocol. We show only one buyer $B$ and three
  sellers $S1$, $S2$, and $S3$. At time 1 the buyer requests bids for
  some good. At time 2 the sellers send their prices for that good. At
  time 3 the buyer picks one of the bids, pays the seller the amount
  and then, at time 4, she receives the good.}
\label{protocol}
\end{center}
\end{figure}

The exchange protocol, seen in Figure~\ref{protocol}, works as
follows: When a buyer $b \in B$ wants to buy a good $g$, she will
advertise this fact. Each seller $s \in S$ that sells that good will
give his bid in the form of a price $p_{s}^{g}$. The buyer will pick
one of these and will pay the seller. All agents will be made aware of
this choice along with the prices offered by all the sellers.. The
winning seller will then return\footnote{In the case of agent/link
  failure, each agent is free to set its own timeouts and assess the
  quality of the never-received good accordingly. Bids that are not
  received in time will, of course, not be considered.} the specified
good. Note that there is no law that forces the seller to return a
good of any particular quality. For example, an agent that sells web
search services returns a set of hits as its good. Each buyer of this
good might determine its quality based on the time it took for the
response to arrive, the number of hits, the relevance of the hits, or
any combination of these and/or other features.  Therefore, it would
usually be impossible to enforce a quality measure that all buyers can
agree with.

It is thus up to the buyer to assess the quality $q$ of the good
received.  Each buyer $b$ also has a value function $V_{b}^{g}(p,q)$
for each good $g \in G$ that she might wish to buy. The function
returns a number that represents the value that $b$ assigns to that
particular good at that particular price and quality. Each seller $s
\in S$, on the other hand, has a cost $c^{g}_{s}$ associated with each
good he can produce.  Since we assume that costs and payments are
expressed in the same units (i.e.  money) then, if seller $s$ gets
paid $p$ for good $g$, his profit will be $\mbox{Profit}(p,c^{g}_{s})
= p - c^{g}_s$.  The buyers, therefore, have the goal of maximizing
the value they get for their transactions, while the sellers have the
goal of maximizing their profits.

\section{Learning recursive models}

Agents placed in the economic society we just described will have to
learn, typically via trial and error, what actions give them the
highest expected reward and under which circumstances. In this section
we will present techniques that these agents might use to maximize
their rewards.

An important question we wish to answer is: when do agents benefit
from having deeper nested models of other agents?  It seems intuitive
that, ignoring computational costs, the agents with more complete
models of others will always do better. We find this to be usually
true; however, while there are instances when it is significantly
better to have deeper models, there are also instances when the
difference is barely noticeable and, instances when its better to
ignore deeper models if they are imperfect or provide no useful
information. These instances are defined in part by the set of other
agents present, their capabilities and preferences, and the dynamics
of the system. In order to precisely determine what these instances
are, and in the hopes of providing a more general framework for
studying the effects of increased agent-modeling capabilities within
our economic model, we have defined a set of techniques that our
agents can use for learning and using models.

We divide the agents into classes that correspond to their modeling
capabilities. The hierarchy we present is inspired by the Recursive
Modeling Method \cite{gmy:96}, but is function-based rather than
matrix-based, and includes learning. We will first describe our agents
at the knowledge level, stating only the type of knowledge the agents
are either trying to acquire through learning, or already have (i.e.
knowledge that was directly implemented by the designers of the
agents), and will then explain the details of how this knowledge was
implemented.

At the most abstract level, we can say that every agent $i$ is trying
to learn the oracle decision function $\Delta_{i}: w \ra a_{i}$, which
maps the state $w$ of the world into the action $a_{i}$ that the agent
should take in that state. This function will not be fixed throughout
the lifetime of the agent because the other agents are also engaged in
some kind of learning themselves.  The agents that try to directly
learn $\Delta_{i}(w)$ we refer to as \textbf{0-level} agents, because
they have no explicit models of other agents.  In fact, they are not
aware that there are other agents in the world.  Any such agent $i$
will learn a decision function \kiwa{} where $w$ is
what agent $i$ knows about its external world and $a_{i}$ is its
rational action in that state. For example, a web search agent might
look at the going price for web searches, in order to determine how
much to charge for its service.

Agents with \textbf{1-level} models of other agents, on the other
hand, are aware that there are other agents out there but have no idea
what the ``interior'' of these agents looks like. They have two kinds
of knowledge--- a set of functions \kijwa{} which capture agent $i$'s
model of each of the other agents ($j$), and \kiwoa{} which captures
$i$'s knowledge of what action to take given $w$ and the collective
actions $\vec{a}_{-i}$ the others will take. We define $\vec{a}_{-i} =
\{a_{1}\cdots a_{i-1},a_{i+1}\cdots a_{n}\}$, where $n$ is the number
of agents. An agent's model of others might not be correct; therefore,
it is not always true that $\delta_{j}(w) = \delta_{ij}(w)$. The
\kijw{} knowledge for all $j \neq i$ turns out to be easier to learn
than the joint action \kiwo{} because the set of possible
hypotheses is smaller.


Agents with \textbf{2-level} models are assumed to have deeper
knowledge about the other agents; that is, they have knowledge of the
form \kijwoa{}.  This knowledge tells them how others determine which
action to take.  They also know what actions others think others are
going to take, i.e. \kijkwa{}, and (like 1-level modelers) what action
they should take given others' actions, \kiwoa{}. Again, the \kijkw{} is
easier to learn that the other two, as long as all agents use the same
features to discriminate among the different worlds (i.e. share the
same $w$).

\subsection{Populating the knowledge}

If the different level agents had to learn all the knowledge then,
since the 0-level agents have a lot less knowledge to learn, they would
learn it much faster. However, in the economic domain, it is likely
that the designer has additional knowledge which could be incorporated
into the agents. The agents we built incorporated extra knowledge
along these lines.

We decided that 0-level agents would learn all their knowledge by
tracking their actions and the rewards they got. These agents,
therefore, receive no extra domain knowledge from the designers and
learn everything from experience. 1-level agents, on the other hand,
have a priori knowledge of what action they should take given the
actions that others will take. That is, while they try to learn
knowledge of the form \kijw{} by observing the actions others take
(i.e. in a form of supervised learning where the other agents act as
tutors), they already have knowledge of the form \kiwo{}. In our
economic domain, it is reasonable to assume that agents have this
knowledge since, in fact, this type of knowledge can be easily
generated. That is, if I know what all the other sellers are going to
bid, and the prices that the buyer is willing to pay, then it is easy
for me to determine which price to bid. We must also point out that in
this domain, the \kiwo{} knowledge cannot be used by a 0-level agent.
If this knowledge had said, for instance, that from some state $w$
agent $i$ will only ever take one of a few possible actions, then this
knowledge could have been used to eliminate impossibilities from the
\kiw{} knowledge of a 0-level agent. However, this situation never
arises in our domain because, as we shall see in the following
Sections, the states used by the agents permit the set of reasonable
actions to always be equal to the set of all possible actions.

The 2-level agents learn their \kijkw{} knowledge from observations of
others' actions, under the already stated assumption that there is
common knowledge of the fact that all agents see the actions taken by
all. The rest of the knowledge, i.e.  \kijwo{} and \kiwo, is built into
the 2-level agents a priori. As with 1-level agents, we cannot use the
\kijwo{} knowledge to add \kijw{} knowledge to a 1-level modeler, because
other agents are also free to take any one of the possible actions in
any state of the world.  There are many reasonable ways to explain how
the 2-level agents came to possess the \kijwo{} knowledge. It
could be, for instance, that the designer assumed that the other
designers would build 1-level agents with the same knowledge we just
described. This type of recursive thinking (i.e.  ``they will do just
as I did, so I must do one better''), along with the obvious
expansion of the knowledge structure, could be used to generate
$n$-level agents, but so far we have concentrated only on the first
three levels.
\begin{table}[p]
\begin{center}
\begin{tabular}{|l||r|l|} \hline
Level   &      Form of Knowledge  & Method of Acquisition\\
\hline \hline
0-level & \kiw{} & Reinforcement Learning\\ \hline
1-level & \kiwo{} & Previously known\\
        & \kijw{}  & Learn from observation   \\ 
\hline 
2-level & \kiwo{} & Previously known \\ 
        & \kijwo{} & Previously known\\ 
        & \kijkw{} & Learn from observation.\\  \hline 
\end{tabular}
\caption{Summary of the forms of knowledge that the different agents
  have or are trying to learn.}
\label{tab:know}
\end{center}
\end{table}
The different forms of knowledge, and their form of acquisition,
are summarized in Table~\ref{tab:know}.  In the following sections, we
talk about each one of these agents in more detail and give some
specifics on their implementation. Our current model emphasizes
transactions over a single good, so each agent is only a buyer or a
seller, but cannot be both.

\subsection{Agents with 0-level models}

Agents with 0-level models must learn everything they know from
observations they make about the environment, and from any rewards
they get. In our economic society this means that buyers see the bids
they receive and the good received after striking a contract, while
sellers see the request for bids and the profit they made (if any).
In general, these agents get some input, take an action, then receive
some reward. This framework is the same framework used in
reinforcement learning, which is why we decided to use a form of
reinforcement learning \cite{sutton:88} \cite{watkins:92}, for
implementing learning in our agents.

Both buyers and sellers will use the equations in the next few sections
for determining what actions to take. But, with a small
probability $\epsilon$ they will choose to explore, instead of
exploit, and will pick their actions at random (except for the fact
that sellers never bid below cost). The value of $\epsilon$ is
initially $1$ but decreases with time to some empirically chosen,
fixed minimum value $\epsilon_{\min}$. That is,
\[
\epsilon_{t+1} = \left\{ \begin{array}{ll} 
    \gamma\epsilon_{t} & \mbox{if $\gamma\epsilon_{t} > \epsilon_{\min}$} \\
    \epsilon_{\min} & \mbox{otherwise}
  \end{array}
\right.
\]

where $0 < \gamma < 1 $ is some annealing factor.
        
\subsubsection{Buyers with 0-level models.}

A buyer $b$ will start by requesting bids for a good $g$. She will
then receive all bids for good $g$ and will pick the seller:
\begin{equation}
  s^{*} = \arg_{s \in S} \max f^{g}(p_{s}^{g})
\end{equation}

This function implements the buyer's $\delta_{b}(w)$ which, in
this case, can be rephrased as $\delta_{b}(p_{1}\ldots p_{|S|})$.
The function $f^{g}(p)$ returns the value the buyer expects to get if
she buys good $g$ at a price of $p$.  It is learned using a simple
form of reinforcement learning, namely:

\begin{equation}
  f^{g}_{t+1}(p) = (1 - \alpha)f^{g}_{t}(p) + \alpha \cdot
  V_{b}^{g}(p,q)
\end{equation}

Here $\alpha$ is the learning rate, $p$ is the price $b$ pays for the
good, and $q$ is the quality she ascribes to it. The learning rate is
initially set to $1$ and, like $\epsilon$, is decreased until it
reaches some fixed minimum value $\alpha_{\min}$.

\subsubsection{Sellers with 0-level models.}
When asked for a bid, the seller $s$ will provide one whose price is
greater than or equal\footnote{We could just as easily have said that
  the price must be strictly greater than the cost.} to the cost for
producing it (i.e. $p^{g}_{s} \ge c^{g}_{s}$). From these prices, he
will chose the one with the highest expected profit:
\begin{equation}
  p^{*}_{s} = \arg_{p \in P} \max h_{s}^{g}(p)
\end{equation}
Again, this function encompasses the seller's $\delta_{s}(g)$
knowledge, where we now have that the states are the goods being sold
$w = g$, and the actions are prices offered $a = p$.  The function
$h_{s}^{g}(p)$ returns the profit $s$ expects to get if he offers good
$g$ at a price $p$. It is also learned using reinforcement learning,
as follows:
\begin{equation}
  h^{g}_{t+1}(p) = (1 - \alpha)h^{g}_{t}(p) + \alpha \cdot
  \mbox{Profit}_{s}^{g}(p)
\end{equation}
where
\begin{equation}
  \mbox{Profit}_{s}^{g}(p) = \left\{ \begin{array}{ll}
      p - c_{s}^{g} & \mbox{if his bid is chosen} \\
      0               &       \mbox{otherwise} \\
        \end{array}
      \right.
\end{equation}

\subsection{Agents with One-level Models}

The next step is for an agent to keep one-level models of the other
agents. This means that it has no idea of what the interior (i.e.
``mental'') processes of the other agents are, but it recognizes the
fact that there are other agents out there whose behaviors influence
its rewards. The agent, therefore, can only model others by looking at
their past behavior and trying to predict, from it, their future
actions. The agent also has knowledge, implemented as functions, that
tells it what action to take, given a probability distribution over
the set of actions that other agents can take.  In the actual
implementation, as shown below, the \kiwo{} knowledge takes into account
the fact that the \kijw{} knowledge is constantly being learned and,
therefore, is not correct with perfect certainty.

\subsubsection{Buyers with one-level models.}
\label{b1l}
A buyer with one-level models can now keep a history of the qualities
she ascribes to the goods returned by each seller. She can, in fact,
remember the last $N$ qualities returned by some seller $s$ for some
good $g$, and define a probability density function $q_{s}^{g}(x)$
over the qualities $x$ returned by $s$ (i.e. $q_{s}^{g}(x)$ returns
the probability that $s$ returns an instance of good $g$ that has
quality $x$). This function provides the $\delta_{bs}(g)$
knowledge.  She can use the expected value of this function to
calculate which seller she expects will give her the highest expected
value.
\begin{eqnarray}
  s^{*} & = &  \arg_{s \in S} \max E
  (V^{g}_{b}(p_{s}^{g},q_{s}^{g}(x))) \nonumber \\
        & = &
  \arg_{s \in S} \max \frac{1}{|Q|} \sum_{x \in Q} q_{s}^{g}(x)\cdot
  V^{g}_{b}(p_{s}^{g},x)
\end{eqnarray}

The $\delta_{b}(g, q_{1} \cdots q_{|S|})$ is given by the previous
function which simply tries to maximize the value the buyer expects
to get.  The buyer does not need to model other buyers since they do
not affect the value she gets.

\subsubsection{Sellers with one-level models.}
\label{s1l}
Each seller will try to predict what bid the other sellers will submit
(based solely on what they have bid in the past), and what bid the
buyer will likely pick. A complete implementation would require the
seller to remember past combinations of buyers, bids and results (i.e.
who was buying, who bid what, and who won). However, it is unrealistic
to expect a seller to remember all this since there are at least
$|P|^{|S|} \cdot |B|$ possible combinations.

However, the seller's one-level behavior can be approximated by having
him remember the last $N$ prices accepted by each buyer $b$ for each
good $g$, and form a probability density function $m_{b}^{g}(x)$,
which returns the probability that $b$ will accept (pick) price $p$
for good $g$. The expected value of this function provides the
$\delta_{sb}(g)$ knowledge. Similarly, the seller remembers
other sellers' last $N$ bids for good $g$ and forms $n_{s}^{g}(y)$,
which gives the probability that $s$ will bid $y$ for good $g$. The
expected value of this function provides the $\delta_{s}(g)$
knowledge.  The seller $s$ can now determine which bid maximizes his
expected profits.

\begin{equation}
p^{*} =  \arg_{p \in P} \max (p - c_{s}^{g}) \cdot
    \prod_{s' \in \{S - s\}} 
        \sum_{p' \in P}
 \left\{ \begin{array}{ll}
                n_{s'}^{g}(p') & \mbox{if $m_{b}^{g}(p') \le
                m_{b}^{g}(p)$} \\
                0               & \mbox{otherwise}
        \end{array}
\right.
\end{equation}

Note that this function also does a small amount of approximation by
assuming that $s$ wins whenever there is a tie\footnote{The complete
  solution would have to consider the probabilities that $s$ ties with
  1, 2, 3,\ldots other agents. In order to do this we would need to
  consider all $|P|^{|S|}$ subsets.}. The function calculates the best
bid by determining, for each possible bid, the product of the profit
and the probability that the agent will get that profit. Since the
profit for lost bids is $0$, we only need to consider the cases where
$s$ wins. The probability that $s$ will win can then be found by
calculating the product of the probabilities that his bid will beat
the bids of each of the other sellers. The function approximates the
$\delta_{s}(g,p_{b},p_{1} \cdots p_{|S|})$ knowledge.

\subsection{Agents with Two-level Models}
The intentional models we use correspond to the functions used by
agents that use one-level models. The agents' \kiwo{} knowledge has
again been expanded to take into account the fact that the deeper
knowledge is learned and might not be correct.  The \kijkw{} knowledge
is learned from observation, under the assumption that there is common
knowledge of the fact that all agents see the bids given by all
agents.

\subsubsection{Buyers with two-level models.}
Since the buyer receives bids from the sellers, there is no need for
her to try to out-guess or predict what the sellers will bid.  She is
also not concerned with what the other buyers are doing since, in our
model, there is an effectively infinite supply of goods. The buyers
are, therefore, not competing with each other and do not need to keep
deeper models of others.

\subsubsection{Sellers with two-level models.}
A seller will model other sellers as if they were using the one-level
models. That is, he thinks they will model others using policy models
and make their decisions using the equations presented in
Section~\ref{s1l}. He will try to predict their bids and then try to
find a bid for himself that the buyer will prefer more than all the
bids of the other sellers. His model of the buyer will also be an
intentional model. He will model the buyers as though they were
implemented as explained in Section~\ref{b1l}. A seller, therefore,
assumes that it has the correct intentional models of other agents.

The algorithm he follows is to first use his models of the sellers to
predict what bids $p_{i}$ they will submit. He has a model of the
buyer $C(s_{1} \cdots s_{n}, p_{1} \cdots p_{n}) \ra s_{i}$, that
tells him which seller she might choose given the set of bids $p_{i}$
submitted by each seller $s_{i}$. The seller $s_{j}$ uses this model
to determine which of his bids will bring him higher profit, by first
finding the set of bids he can make that will win:
\begin{equation}
  P' = \{p_{j} | p_{j} \in P, s_{j} = C( s_{1} \cdots s_{j} \cdots s_{n},
  p_{1} \cdots p_{j} \cdots p_{n}) \}
\end{equation}

And from these finding the one with the highest profit:
\begin{equation}
  p^{*} = \arg_{p \in P'} \max (p - c_{s}^{g})
\end{equation}
These functions provide the $\delta_{s}(g,p_{b},p_{1} \cdots
p_{|S|})$ knowledge.

\section{Tests}
\label{tests}

Since there is no obvious way to analytically determine how different
populations of agents would interact and, of greater interest to us,
how much better (or worse) the agents with deeper models would fare,
we decided to implement a society of the agents described above and
ran it to test our hypotheses. In all tests, we had $5$ buyers and $8$
sellers. The buyers had the same value function $V_{b}(p,q) = 3q - p$,
which means that if $p = q$ then the buyers will prefer the seller
that offers the higher quality. The quality that they perceived was
the same only on average, i.e. any particular good might be thought to
have quality that is slightly higher or lower than expected. All
sellers had costs equal to the quality they returned in order to
support the common sense assumption that quality goods cost more to
produce. A set of these buyers and sellers is what we call a
\emph{population}. We tried various populations; within each
population we kept constant the agents' modeling levels, the value
assessment functions and the qualities returned.  The tests involved a
series of such populations, each one with agents of different modeling
levels, and/or sellers with different quality/costs.  We also set
$\alpha_{\min} = .1$, $\epsilon_{\min} = .05$, and $\gamma = .99$.
There were $100$ runs done for each population of agents, each run
consisting of $10000$ auctions (i.e.  iterations of the protocol). The
lessons presented in the next section are based on the averages of
these $100$ runs.

\section{Lessons}
\label{lessons}
From our tests we were able to discern several lessons about the
dynamics of different populations of agents. Some of these lead to
methods that can be used to make quantitative predictions about
agents' performance, while others make qualitative assessments about
the type of behaviors we might expect. We detail these in the next
subsections, and summarize them in Table~\ref{table1}.

\subsection{Micro versus macro behaviors.}
In all tests, we found the behavior for any particular run does not
necessarily reflect the average behavior of the system. The prices
have a tendency to sometimes reach temporary stable points. These
\emph{conjectural equilibria}, as described in \cite{hu:96}, are
instances when all of the agents' models are correctly predicting the
others' behavior and, therefore, the agents do not need to change
their models or their actions. These conjectural equilibria points are
seldom global optima for the agents. If one of our agents finds itself
at one of these equilibrium points, since the agent is
always exploring with probability $\epsilon$, it will in time discover
that this point is only a local optima (i.e. it can get more profit
selling/buying at a different price) and will change its actions
accordingly. Only when the price is an equilibrium price\footnote{That
  is, $p$ is an equilibrium price if every seller that can sell at
  that price (i.e. those whose cost is less than $p$) does.} do we
find that the agents continue to forever take the same actions,
leaving the price at its equilibrium point.

In order to understand the more significant macro-level behaviors of
the system, we present results that are based on the averages from
many runs. While these averages seem very stable, and a good first
step in learning to understand these systems, in the future we will
need to address some of the micro-level issues. We do notice from our
data that the micro-level behaviors (e.g. temporary conjectural
equilibria, price fluctuations) are much more closely tied, usually
in intuitive ways, to the agents' learning rate $\alpha$ and
exploration rate $\epsilon$.  That is, higher rates for both of these
lead to more price fluctuations and shorter temporary equilibria.

\subsection{0-level buyers and sellers.}
\begin{figure*}[p]
  \chartll{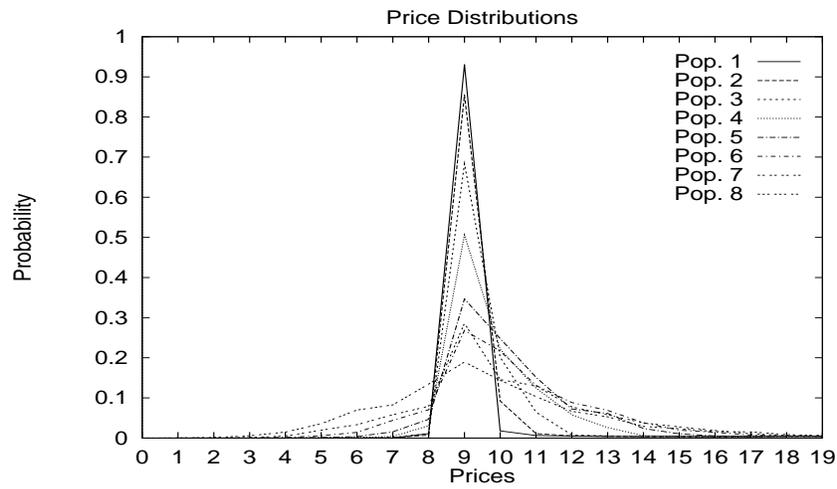}
\caption{Price distributions for populations of 0-level buyers and
  sellers. The prices are $0 \cdots 19$. The columns represent the
  percentage of time the good was sold at each price, in each
  population. In {\em p1} sellers return qualities
  $\{8,8,8,8,8,8,8,8\}$, in {\em p2} its $\{8,8,8,8,8,8,7,8\}$, and so
  on such that by {\em p8} its $\{1,2,3,4,5,6,7,8\}$. The highest peak
  in all populations corresponds to price $9$.}
\label{t12.pd}
\end{figure*}
\begin{figure}[p]
  \begin{center}
    \leavevmode
    \chartl{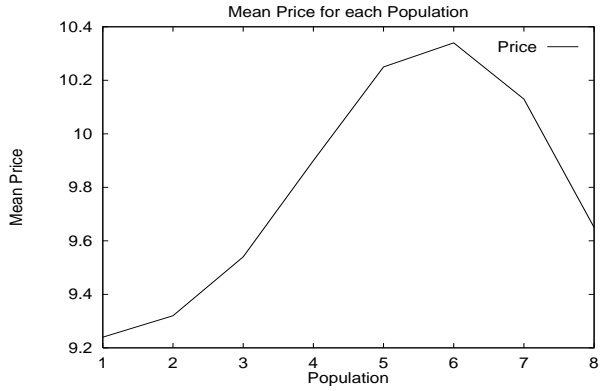}
    \caption{Mean price for the populations from Figure~\ref{t12.pd}}
    \label{t12.mean}
  \end{center}
\end{figure}

This type of population is equivalent to a ``blind'' auction, where
the agents only see the price and the good, but are prevented from
seeing who the seller (or buyer) was. As expected, we found that an
equilibrium is reached as long as all the sellers are providing the
same quality. This is the case for population~1 in
Figure~\ref{t12.pd}. Otherwise, if the sellers offer different quality
goods, the price fluctuates as the buyers try to find the price that
on the whole returns the best quality, and the sellers try to find the
price\footnote{Remember, the sellers are constrained to return a fixed
  quality. They can only change the price they charge.} the buyers
favor. In these populations, the sellers offering the higher quality,
at a higher cost, lose money.  Meanwhile, sellers offering lower
quality, at a lower cost, earn some extra income by selling their low
quality goods to buyers that expect, and are paying for, higher
quality. As more sellers start to offer lower quality, we find that
the mean price actually {\em increases}, evidently because price acts
as a signal for quality and the added uncertainty makes the higher
prices more likely to give the buyer a higher value. We see this in
Figure~\ref{t12.pd}, where population~1 has all sellers returning the
same quality while in each successive population more agents offer
lower quality.  The price distribution for population~1 is
concentrated on 9, but for populations 2 through 6 it flattens and
shifts to the right, increasing the mean price. It is only by
population~7 when it starts to shift back to the left, thus reducing
the mean price, as seen in Figure~\ref{t12.mean}. That is, it is only
after a significant number of sellers start to offer lower quality
that we see the mean price decrease.

\subsection{0-level buyers and sellers, plus one 1-level seller.}
\label{0lb01ls}
\begin{figure*}[p]
  \chartll{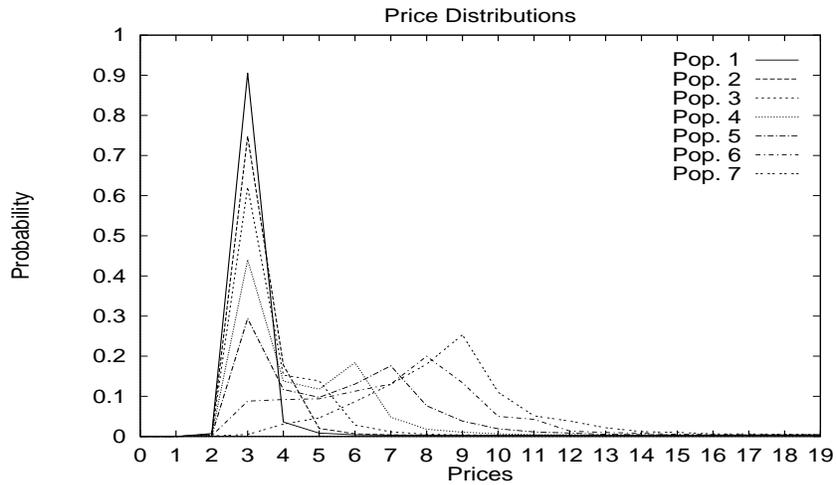}
\caption{Price distributions for populations of 0-level buyers and
  0-level sellers plus one 1-level seller. In population~1 sellers
  return qualities $\{2,2,2,2,2,2,2,2\}$, in pop. $2$ its
  $\{2,3,2,2,2,2,2,2\}$, and so on such that by pop. $7$ it's
  $\{2,3,4,5,6,7,8,2\}$. The 1-level seller is the last on the
  list. It always returns quality 2.}
\label{t3.pd}
\end{figure*}
In these population sets we explored the advantages that a 1-level
seller has over identical 0-level sellers. The advantage was
non-existent when all sellers returned the same quality (i.e. when the
prices reached an equilibrium as shown in population~1 in
Figure~\ref{t3.pd}), but increased as the sellers started to diverge
in the quality they returned. In order to make these findings useful
when building agents, we needed a way to make quantitative predictions
as to the benefits of keeping 1-level models.  It turns out that these
benefits can be predicted, not by the population type as we had first
guessed, but by the price volatility.
\begin{figure}[p]
  \chart{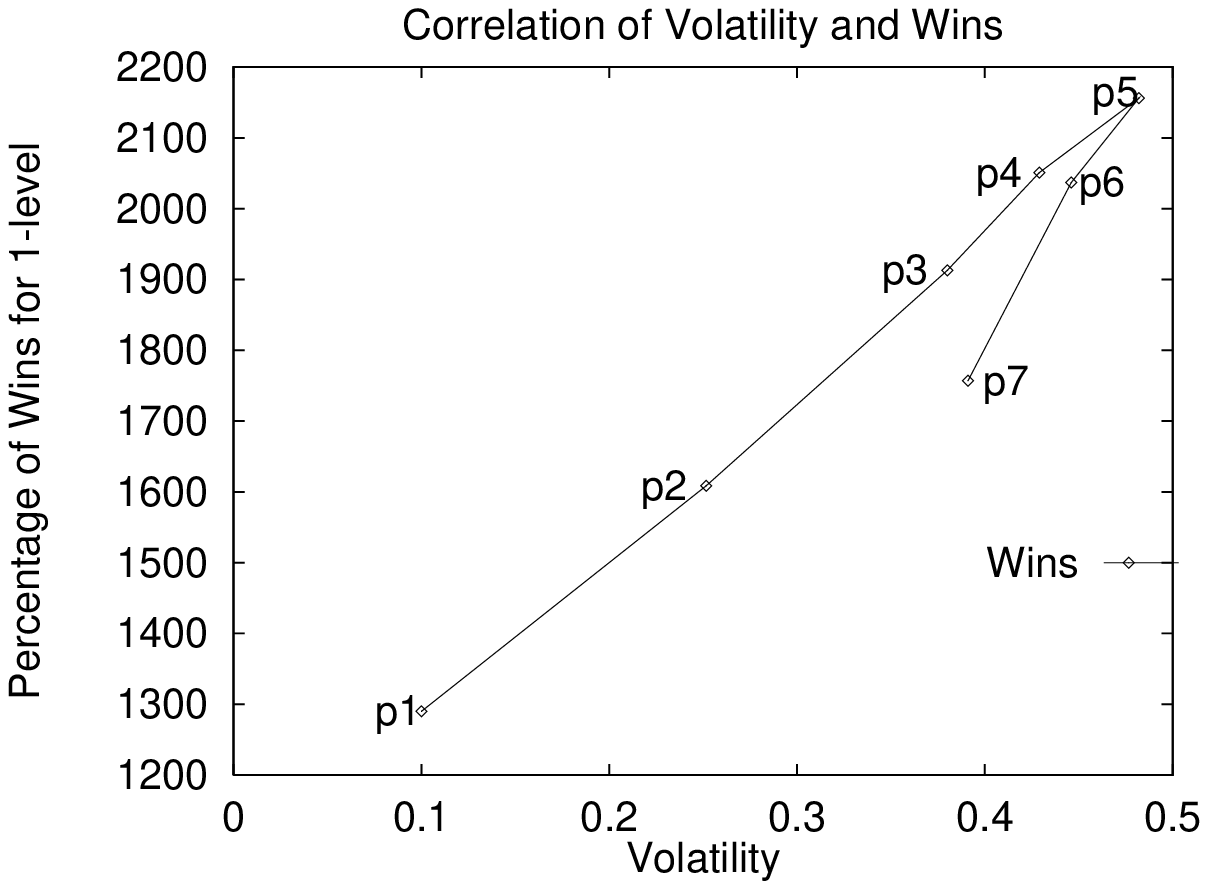}
\caption{Scatter plot of volatility versus the percentage of time that
  the 1-level seller wins ($w$). The populations are the same as in
  Figure~\ref{t3.pd}.} 
\label{t3.av}
\end{figure}
We define {\em volatility} as the number of times the price changes
from one auction to the next, divided by the total number of auctions.
Figure~\ref{t3.av} shows the linear relation between volatility and
the percentage of times the 1-level seller wins. The two lines
correspond to two ``types'' of volatility. The first line includes
populations 1 through 5 ({\em p1}-{\em p5}). It reflects the case
where the buyers' second-favorite (and possibly, the third, fourth,
etc.) equilibrium price is greater than her most preferred price. In
these cases the buyers and sellers fight among the two most preferred
prices, the sellers pulling towards the higher equilibrium price and
the buyers towards the lower one, as shown by the two peaks in
populations 4 and 5 in Figure~\ref{t3.pd}. The other line, which
includes populations 6 and 7, corresponds to cases where the
buyers' preferred equilibrium price is greater than the runner-ups. In
these cases there is no contest between two equilibria. We observe
only one peak in the price distribution for these populations.

The slope of these lines can be easily calculated and the resulting
function can be used by a seller agent for making a {\em quantitative}
prediction as to how much he would benefit by switching to 1-level
models. That is, he could measure price volatility, multiply it by the
appropriate slope, and the resulting number would be the percentage of
times he would win. However, for this to work the agent needs to know
that all eight buyers and five sellers are 0-level modelers because
different types of populations lead to different slopes. Also,
slight changes in our learning parameters ($.02 \leq \epsilon_{\min}
\leq .08$ and $.05 \leq \alpha_{\min} \leq .2$) lead to slight changes
in the slopes so these would have to be taken into account if the
agent is actively changing its parameters.

We also want to make clear a small caveat, which is that the
volatility that is correlated to the usefulness of keeping 1-level
models is the volatility of the system {\em with} the agent already
doing 1-level modeling. Fortunately, our experiments show that having
one agent change from 0-level to 1-level does not have a great effect
on the volatility as long as there are enough (i.e. more than five or
so) other sellers.

The reason volatility is such a good predictor is that it serves as an
accurate assessment of how dynamic the system is and, in turn, of the
complexity of the learning problem faced by the agents. It turns out
that the learning problem faced by 1-level agents is ``simpler'' than
the one faced by 0-level modelers.  Our 0-level agents use
reinforcement learning to learn a good match between world states and
the actions they should take. The 1-level agents, on the other hand,
can see the actions other agents take and do not need to learn their
models through indirect reinforcements. They instead use a form of
supervised learning to learn the models of others. Since 1-level
agents need fewer interactions to learn a correct model, their models
will, in general, be better than those of 0-level agents in direct
proportion to the speed with which the target function changes. That
is, in a slow-changing world both of them will have time enough to
arrive at approximately correct models, while in a fast-changing world
only the 1-level agents will have time to arrive at an approximately
correct model. This explains why high price volatility is correlated
to an increase in the 1-level agent's performance. However, as we saw,
the relative advantages for different volatilities (i.e. the slope in
Figure~\ref{t3.av}) will also depend on the shape of the price
distribution and the particular population of agents.

Finally, in all populations where the buyers are 0-level, we saw that
it really pays for the sellers to have low costs because this allows
them to lower their prices to fit almost any demand. Since the buyers
have 0-level models, the sellers with low quality and cost can raise
their prices when appropriate, in effect ``pretending'' to be the
high-quality sellers, and make an even more substantial profit. This
extra profit comes at the cost of a reduction in the average value
that the buyers receive. In other words, the buyers get less value
because they are only 0-level agents and are less able to detect the
sellers' deception. In the next Section we will see how this is not
true for 1-level buyers.
\begin{figure}[p]
    \leavevmode
    \chartl{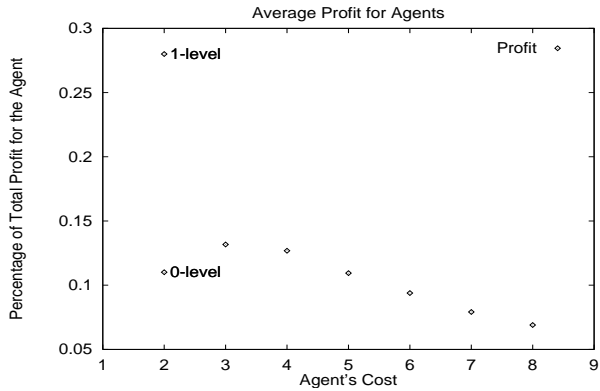}
    \caption{Agents' profit in population~7 from Figure~\ref{t3.pd}.
      The sellers return qualities of $\{2,3,4,5,6,7,8,2\}$,
      respectively, where the 1-level seller is the first on the
      list.}
    \label{t3.profit}
\end{figure}

Of course, the 1-level sellers were more successful at this deception
strategy than the 0-level sellers.  Figure~\ref{t3.profit} shows the
profit of several agents in a population as a function of their cost.
We can see how the 0-level agents' profit decreases with increasing
costs, and how the 1-level agent's profit is much higher than the
0-level with the same costs. We also notice that, since the 0-level
agents are not as successful as the 1-level at taking advantage of
their low costs, the first 0-level seller (that returns quality 2) has
lower profit than the rest as some of his profit was taken away by the
1-level seller (that returns the same quality).

\subsection{1-level buyers and 0 and 1-level sellers.}
\begin{figure}[p]
    \leavevmode
    \chartl{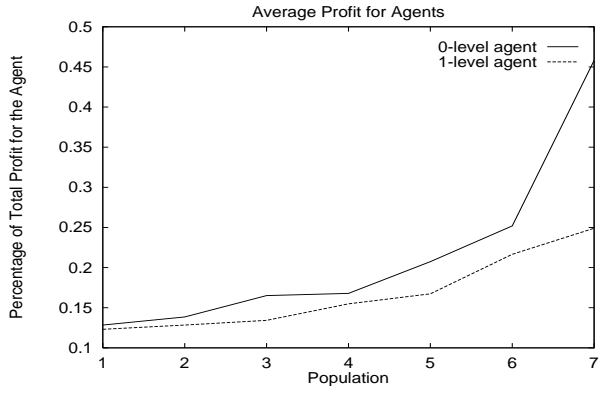}
    \caption{Profit of a 1-level seller and a 0-level seller, both
      with the same costs and quality, in a population with 1-level
      buyers.}
    \label{t21.profit}
\end{figure}
In these populations the buyers have the upper hand. They quickly
identify those sellers that provide the highest quality goods and buy
exclusively from them. The sellers do not benefit from having deeper
models; in fact, Figure~\ref{t21.profit} shows how the 1-level
seller's profit is less than that of a similar 0-level seller because
the 1-level seller tries to charge higher prices than the 0-level
seller. The 1-level buyers do not fall for this trick--- they know
what quality to expect, and buy more from the lower-priced 0-level
seller(s). We have here a case of erroneous models--- 1-level sellers
assume that buyers are 0-level, and since this is not true, their
erroneous deductions lead them to make bad decisions. To stay a step
ahead, sellers would need to be 2-level in this case.

In Figure~\ref{t21.profit}, the first population has all sellers
returning a quality of 8 while by population~7 they are returning
qualities of $\{8,2,3,4,5,6,7,8\}$, respectively, with the 1-level
always returning quality of 8. We notice that the difference in
profits between the 0-level and the 1-level increases with successive
populations. This is explained by the fact that in the first
population all seven 0-level sellers are returning the same quality,
while by population~7 only the 0-level pictured (i.e. the first one)
is still returning quality 8. This means that his competition, in the
form of other 0-level sellers returning the same quality, decreases
for successive populations. Meanwhile, in all populations there is
only one 1-level seller who has no competition from other 1-level
sellers. To summarize, the 0-level seller's profit is always higher
than the similar 1-level seller's, and the difference increases as
there are fewer other competing 0-level sellers who offer the same
quality.

\subsection{1-level buyers and several 1-level sellers.}
\begin{figure}[p]
  \chartl{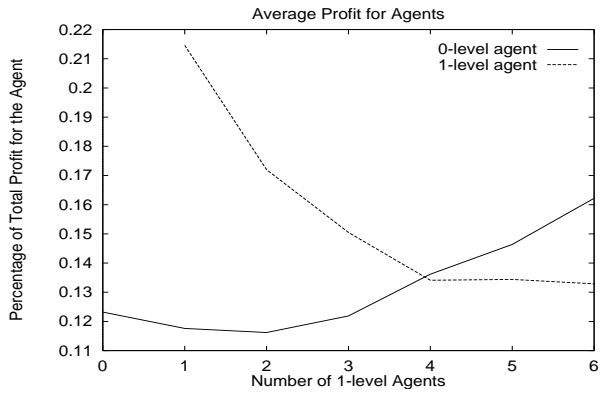}
  \caption{Profit of 1-level seller and similar 0-level seller as a
    function of the number of 1-level sellers in the population. There
    are a total of $8$ agents in all populations.}
  \label{t23.profit}
\end{figure}
We have shown how 1-level sellers do better, on average, than 0-level
sellers when faced with 0-level buyers, but this is not true anymore
if too many 0-level sellers decide to become 1-level.
Figure~\ref{t23.profit} shows how the profits of a 1-level seller
decrease as he is joined by other 1-level sellers. In this Figure the
sellers are returning qualities of $\{2,2,2,2,2,3,4\}$. Initially they
are all 0-level, then one of the sellers with quality 2 becomes
1-level (he is the seller shown in the Figure), then another one and
so on\ldots until there is only one 0-level seller with quality two.
Then the seller with quality three becomes 1-level and, finally the
seller with quality four becomes 1-level. At this point we have six
1-level sellers and one 0-level seller.  We can see that with more
than four 1-level sellers the 0-level seller is actually making more
profit than the similar 1-level seller. The 1-level seller's profit
decreases because, as more sellers change from 0 to 1-level, they are
competing directly with him since they are offering the same quality
and are the same level. Notice that the 1-level seller's curve
flattens after four 1-level sellers are present in the population.
The reason is that the next sellers to change over to 1-level return
qualities of 3 and 4, respectively, so that they do not compete
directly with the seller pictured. His profits, therefore, do not keep
decreasing. 

For this test, and other similar tests, we had to use a population of
sellers that produce different qualities because, as explained in
Section~\ref{0lb01ls}, if they had returned the same quality then an
equilibrium would have been reached which would prevent the 1-level
sellers from making a significantly greater profit than the 0-level
sellers.

\subsection{1-level buyers and 1 and 2-level sellers.}
Assuming that the 2-level seller has perfect models of the other
agents, we find that he wins an overwhelming percentage of the time.
This is true, surprisingly enough, even when some of the 1-level sellers
offer slightly higher quality goods. However, when the quality
difference becomes too great (i.e. greater than 1), the buyers finally
start to buy from the high quality 1-level sellers. This case is very
similar to the ones with 0-level buyers and 0 and 1-level sellers and
we can start to discern a recurring pattern. In this case, however, it
is much more computationally expensive to maintain 2-level models. On
the other hand, since these 2-level models are perfect, they are
better predictors than the 1-level, which explains why the 2-level
seller wins much more than the 1-level seller from
Section~\ref{0lb01ls}. 

\section{Conclusions}
\begin{table*}[p]
\begin{center}
\begin{tabular}{|l|l||l|} \hline 
  Buyers & Sellers & Lessons \\ \hline \hline 0-level & 0-level
  &Equilibrium reached only when all sellers
  offer \\
  & &the same quality. Otherwise, we get oscillations. \\ 
  & &Mean price increases when quality offered decreases. \\ \hline
  0-level & Any & Sellers have big incentives to lower quality/cost. \\ \hline
  0-level & 0-level and & 1-level seller beats others. \\
  &one 1-level & Quantitative advantage of being
  1-level predicted by \\
  & & volatility and price distribution. \\ \hline
  0-level & 0-level and   & 1-level sellers do better, as long as there \\
  & many 1-level & are not too many of them. \\ \hline
  1-level & 0-level and   & Buyers have upper hand. They buy from the most \\
  & one 1-level   & preferred seller. \\
  & & 1-level sellers are usually at a disadvantage.\\ \hline 
  1-level &  1-level and & Since 2-level has perfect models, it wins
  an \\
  & one 2-level   & overwhelming percentage of time, except \\
  & & when it offers a rather lower quality. \\ \hline
\end{tabular}
\caption{\textbf{Summary of lessons}. In all cases the buyers had identical
  value and quality assessment functions. Sellers were constrained to
  always return the same quality.}
\label{table1}
\end{center}
\end{table*}
We have presented a framework for the development of agents with
incremental modeling/learning capabilities, in an economic society of
agents. These agents were built, and the execution of different agent
populations leads us to the discovery of the lessons summarized in
Table~\ref{table1}. The discovery of volatility and price
distributions as predictors of the benefits of deeper models will be
very useful as guides for deciding how much modeling capability to
build into an agent. This decision could either be done prior to
development or, given enough information, it could be done at runtime.
We are also encouraged by the fact that increasing the agents'
capabilities changes the system in ways that we can recognize from our
everyday economic experience.

Some of the agent structures shown in this paper are already being
implemented into the UMDL \cite{atkins:96}. We have a basic economic
infrastructure that allows agents to engage in commerce, and the
agents use customizable heuristics for determining their strategic
behavior. We are working on incorporating the more advanced modeling
capabilities into our agents in order to enable more interesting
strategic behaviors.

Our results showed how sellers with deeper models fare better, in
general, even when they produce less valuable goods. This means that
we should expect those types of agents to, eventually, be added into
the UMDL\footnote{If not by us, then by a profit-conscious third
  party.}. Fortunately, this advantage is diminished by having buyers
keep deeper models. We expect that there will be a level at which the
gains and costs associated with keeping deeper models balance out for
each agent. Our hope is to provide a mechanism for agents to
dynamically determine this cutoff and constantly adjust their behavior
to maximize their expected profits given the current system behavior.
The lessons in this paper are a significant step in this
direction. We have seen that one  needs to look at price
volatility and at the modeling levels of the other agents to determine
what modeling level will give the highest profits. We have also
learned how buyers and sellers of different levels and offering
different qualities lead to different system dynamics which, in turn,
dictate whether the learning of nested models is useful or not.

We are considering the expansion of the model with the possible
additions of agents that can both buy and sell, and sellers that can
return different quality goods. Allowing sellers to change the quality
returned to fit the buyer will make them more competitive against
1-level buyers. We are also continuing tests on many different types
of agent populations in the hopes of getting a better understanding of
how well different agents fare in the different populations.

In the long run, another offshoot of this research could be a better
characterization of the types of environments and how they
allow/inhibit ``cheating'' behavior in different agent populations.
That is, we saw how, in our economic model, agents are sometimes
rewarded for behavior that does not seem to be good for the community
as a whole (e.g. when some of the sellers raised their price while
lowering the quality they offered). The rewards, we are finding, start
to diminish as the other agents become ``smarter''.  We can intuit
that the agents in these systems will eventually settle on some level
of nesting that balances their costs of keeping nested models with
their gains from taking better actions \cite{kauffman:94}. It would be
very useful to characterize the environments, agent populations, and
types of ``equilibria'' that these might lead to, especially as
interest in multi-agent systems grows.

\nocite{durfee:94a, akerlof:70,watkins:92,
  artificial:societies,tambe:96, russell:95, mullen:96, vidal:96a,
  vidal:96c} 

\bibliographystyle{apalike} 

\bibliography{jetai-bib}

\end{document}